\documentclass[%
reprint, 
superscriptaddress,
nobibnotes,
amsmath,amssymb,
aps,
floatfix,
]{revtex4-2}
\usepackage[utf8]{inputenc}

\usepackage[british]{babel}
\usepackage{csquotes}
\usepackage[expansion=false, babel=true, protrusion = true]{microtype}

\usepackage{hyperref}

\usepackage{mathtools}
\usepackage{graphicx}

\usepackage{physics}

\usepackage[%
]{todonotes}
\let\oldtodo\todo
\renewcommand{\todo}[1]{\oldtodo[inline,color=red!40]{#1}}

\newcounter{myfigpanel}[figure]
\newcounter{myfigpanelonly}[figure]

\newcommand{\panelletter}[1]{\refstepcounter{myfigpanel}\label{#1}\refstepcounter{myfigpanelonly}\label{onlyletter:#1}\alph{myfigpanel}}
\newcommand{\panel}[1]{(\protect\panelletter{#1})}
\let\origcaption\caption
\let\caption\undefined
\DeclareRobustCommand{\caption}[1]{\origcaption{\protect\setcounter{myfigpanel}{0}\protect\setcounter{myfigpanelonly}{0}#1}}

\newcommand\cs{~}
\newcommand\InPartS{\mbox{InPartS}}
\renewcommand\vec[1]{\mathbf{#1}}
\newcommand*\uvec[1]{\hat{\vec{#1}}}
\newcommand*\tens[1]{\underline{\boldsymbol{\mathrm{#1}}}}

\newcommand{\hypergeo}[1]{\ensuremath{{}_2F_1\left(\textstyle{1, \frac{\mu}{\alpha}; 1 + \frac{\mu}{\alpha};1 - \frac{#1^2}{R_0^2}}}\right)}

\newcommand*\ust{\tens u^\text{ST}}
\newcommand*\lmax{l^\text{max}}
\newcommand*{\rR}[2][]{\left(\frac{r}{R_0}\right)^{#1\mathllap{#2}}}
\newcommand*{\Rr}[2][\hphantom{-}]{\rR[#1]{-#2}}

\newcommand{\suppfigFEFields}{Supp.~Fig.~S2}
\newcommand{\suppfigPolySnaps}{Supp.~Fig.~S3}
\newcommand{\suppfigFitLandscape}{Supp.~Fig.~S4}
\newcommand{\suppfigHockeyStick}{Supp.~Fig.~S5}

\newcommand\MPIDS{\affiliation{Max Planck Institute for Dynamics and Self-Organization, Göttingen, Germany}}
\newcommand\UGOE{\affiliation{University of Göttingen, Institute for Dynamics of Complex Systems, Göttingen, Germany}}
\newcommand{\RPCTP}{\affiliation{Rudolf Peierls Centre for Theoretical Physics, University of Oxford, Oxford OX1 3PU, United Kingdom}}

\begin{document}

\title{Prediction and control of geometry-induced nematic order in growing multicellular systems}
\author{Lukas Hupe}
\MPIDS\UGOE
\author{Jonas Isensee}
\MPIDS\UGOE
\author{Ramin Golestanian}
\MPIDS\UGOE\RPCTP
\author{Philip Bittihn}
\email{philip.bittihn@ds.mpg.de}
\MPIDS\UGOE

\begin{abstract}
In densely-packed two-dimensional systems of growing cells, such as rod-shaped bacteria, a number of experimental and numerical studies report distinct patterns of nematic orientational order in the presence of confinement.
So far, these effects have been explained using variations of growing active nematic continuum theories, which incorporate feedback between growth-induced active stresses, the resulting material flow and nematic orientation, and were adapted to the specific geometry under investigation.
Here, we first show that a direct, analytical prediction of orientation patterns based on a simple isotropic-growth assumption and the shear rate tensor of the expansion flow already covers previously observed cases.
We use this method to tune orientation patterns and net topological defect charge in a systematic way using domain geometry, confirmed by agent-based simulations.
We then show how this framework can be extended to quantitatively capture alignment strength, and explore its potential for cross-prediction across different geometries.
Our simplified and unifying theoretical framework highlights the role of domain geometry in shaping nematic order of growing systems, and thereby provides a way to forward-engineer desired orientation patterns.
\end{abstract}

\maketitle
\section{Introduction}

Multicellular systems such as bacterial colonies and epithelial tissues feature intricate, dynamic patterns that range from swirling, turbulent flows\cs\cite{alert_active_2022} to topological defects in cell orientation\cs\cite{saw_topological_2017, copenhagen_topological_2020, endresen_topological_2021}.
These collective behaviors are not only of fundamental interest in the field of active matter but are also relevant for biological processes such as morphogenesis~\cs\cite{doostmohammadi_defectmediated_2016,hoffmann_defectmediated_2022} and wound healing~\cs\cite{basan_alignment_2013,tetley_fluidity_2019}.
A central driver of this complexity is the ability of cells to grow and divide, an inherently out-of-equilibrium process that distinguishes these systems from classical condensed matter\cs\cite{hallatschek_proliferating_2023}.
Recent studies have increasingly explored how this intrinsic activity influences large-scale organization, leading to emergent flows, formation of complex pattern\cs\cite{trejo_elasticity_2013,yan_mechanical_2019,nijjer_mechanical_2021}, nematic alignment\cs\cite{volfson_biomechanical_2008,zhang_morphogenesis_2021}, and transitions between fluid-like and solid-like behavior\cs\cite{ranft_fluidization_2010,basan_dissipative_2011,mao_transitions_2024}.

A particularly striking manifestation of this activity can be observed in systems composed of elongated, rod-shaped cells that grow primarily along their long axis before division.
When constrained to grow on planar surfaces such as agar, or within microfluidic chambers, non-motile rods can form dense, quasi-two-dimensional monolayers.
In these systems, steric interactions between the anisotropic cells induce nematic alignment, with the degree and orientation of alignment strongly effected by the system geometry and boundary conditions:
In a commonly studied example, growth in confined channels promotes global alignment along the channel axis\cs\cite{volfson_biomechanical_2008, orozco-fuentes_order_2013, you_confinementinduced_2021, isensee_stress_2022}, while unconfined expansion leads to fluctuating microdomains without long-range order\cs\cite{dellarciprete_growing_2018, you_geometry_2018, isensee_sensitive_2025}.
Other geometries, such as inward radial growth or curvature-induced strain, can give rise to radial alignment and topologically rich orientational fields\cs\cite{basaran_largescale_2022, langeslay_strain_2024}.

To describe these alignment patterns on a continuum level, many authors have turned to nematic liquid crystal theory, and in particular, the Landau-de Gennes framework\cs\cite{degennes_physics_1993}.
This formalism captures orientational order through the $\tens{Q}$-tensor, a symmetric, traceless field representing both the direction and magnitude of local alignment.
The dynamics of this tensor are derived from a free energy functional that includes bulk alignment terms and elastic penalties for spatial distortions.
In active nematics, such as those driven by motility or internal force generation, this approach has been successfully used to describe the spontaneous creation and annihilation of topological defects and how local force generation can create to novel flow instabilities leading to active turbulence\cs\cite{marchetti_hydrodynamics_2013,thampi_instabilities_2014,blow_biphasic_2014,doostmohammadi_defectmediated_2016,dellarciprete_growing_2018,alert_universal_2020}.

In the case of dense growing colonies dominated by steric interactions, the colony itself acts as the fluid in which individual particles are suspended.
In particular, many microscopic models of these systems are set in the dry limit and completely lack hydrodynamic interactions\cs\cite{wensink_mesoscale_2012, basan_alignment_2013}.
Several works have used this idea to connect observed global alignment patterns to the structure of the growth-generated flow field\cs\cite{dellarciprete_growing_2018, nijjer_mechanical_2021, basaran_largescale_2022}.
For instance, Langeslay and Juarez\cs\cite{langeslay_strain_2024} showed that the local shear rate experienced by an elongating rod can predict its preferred orientation.
In this view, alignment is not governed by long-range hydrodynamics but rather by local kinematics---shear alignment arising from growth-induced deformations is sufficient to explain the emergence of ordered patterns in many geometries.

Here, we build on this understanding to develop a general framework that predicts large-scale orientational order in growing rod systems based solely on the geometry and boundary conditions of the colony.
By introducing an \emph{isotropic-growth assumption}, we suppress the feedback that would modify the material flow depending on local alignment of particles, effectively making flow-generation independent of the current particle orientation and resulting in a strictly unidirectional coupling: from the material flow to orientation dynamics.
This simplification allows us to analytically compute preferred particle orientations from the geometry of the system alone, via the shear component of the emerging flow field.
With this approximation, we recover alignment patterns in a range of previously studied radially symmetric systems, including isotropic expansion, radial outward and inward growth, channel geometries, and curvature-induced strain\cs\cite{dellarciprete_growing_2018, you_geometry_2018,you_confinementinduced_2021,basaran_largescale_2022, langeslay_strain_2024}, highlighting the universality of shear alignment in growth-driven systems.
We demonstrate the power of this approach on a new class of example systems, polygonal domains with absorbing boundaries.
These systems exhibit highly heterogeneous global orientation patterns and distinct topological charge of the nematic director, which we verify numerically in particle-based simulations.
Finally, we introduce a minimal extension of the framework to incorporate the finite time scales of both reorientation and decay of nematic order, still based on the isotropic-growth assumption.
This dynamic correction improves quantitative agreement with simulations while preserving analytical tractability. Taken together, our results offer a unifying, geometry-based perspective on orientational order in growing active matter systems.

\section{Shear alignment due to isotropic growth}
\subsection{Theory for radial systems}
\label{sec:radialsystems_theo}
To derive a prediction for orientation patterns from system geometry alone, we assume that a colony of growing rods, expanding along their long axis and interacting only through steric repulsion, behaves like an incompressible expanding nematic fluid.
To describe the orientations of the individual rods as a coarse-grained field, we use the nematic tensor $\tens{Q} = q (\uvec n \otimes\uvec n - \tens I/2)$, a traceless symmetric second order tensor that encodes local alignment strength $q$ and preferred orientation $\uvec n$.

For simplicity, we restrict our considerations to systems in which particle density, flow and orientation have reached a steady state.
Practically, this requires some way of balancing the ongoing addition of cellular material: This could happen either via local cell removal akin to tissue homeostasis\cs\cite{pollack_competitive_2022} or via absorbing boundary conditions which, in a real system, could correspond to boundaries of a confined space beyond which cells exit into open space and are washed away (which can also be realized in microfluidic devices\cite{bittihn_genetically_2020}).
Here, we focus on the latter case, thus leaving growth \emph{locally unbalanced} and causing strong expansion flow fields that advect matter from the bulk to the boundaries.

In the literature, the evolution of the nematic tensor and its coupling to a flow field is usually described with an equation of the form (compare  \cite{volfson_biomechanical_2008, doostmohammadi_defectmediated_2016, srivastava_negative_2016, dellarciprete_growing_2018})
\begin{equation}
\label{eq:basic_q}
\frac{\text{D}}{\text{D}t}\tens{Q} = %
\beta\ust + \tens\omega\tens{Q} - \tens{Q}\tens\omega + \mathcal{O}(\tens{Q})
\end{equation}
where $\text{D}/\text{D}t$ is the co-moving derivative $(\partial_t + \vec v\cdot\nabla)$.
The terms $\ust = (\partial_i v_j + \partial_j v_i)/2 - \frac{\delta_{ij}}{2}\partial_k v_k$ and $\tens{\boldsymbol\omega} = (\partial_i v_j - \partial_j v_i)/2$ are the (traceless) symmetric and antisymmetric components of the velocity gradient tensor $\partial_i v_j$, respectively, representing the shear rate and rotation of the flow field.
The higher order terms in $\tens{}Q$ (and $\nabla\tens{}Q$) are usually in the form of a so-called molecular field, derived from elastic and nematic free energy considerations.

In this equation, the only term that is not proportional to $\tens Q$---and therefore the only term that can drive an isotropic system into an anisotropic state---is the shear alignment term $\beta \ust$.
It is thus reasonable to assume that in the regime of small $\tens Q$, where all other terms in Eq.~\eqref{eq:basic_q} remain small, the initial anisotropic bias given by $\ust$ determines the preferred alignment directions profile.

However, here, unlike in many other active nematic systems, there is no externally imposed flow field. Instead, the rods are advected and reoriented by the collective flow generated through their own expansion and steric repulsion.
Since, for growing rods, the local expansion processes are parallel to the nematic director, conventional descriptions of these systems require dealing with strong feedback between $\tens{Q}$ and the flow field $\vec{v}$, mediated e.g.\ by an anisotropic active stress term.
Here, we assume that the flow field can be approximated using only isotropic growth, and a scalar pressure $p$, cutting the feedback and enabling us to directly compute the expansion flow field.
Starting with a continuity equation for an incompressible fluid, we can derive
\begin{align}
\div\vec{v} &= \alpha\label{eq:velocity_divergence}\quad\text{and equivalently}\\
\nabla^2 p &= -\frac{\alpha}{\zeta}\,,
\label{eq:pressure_poisson}
\end{align}
where $\alpha$ is the (local) volume expansion rate, acting as the only source term, and where velocity $\vec{v}$ and pressure $p$ are linked by the overdamped equation of motion $\vec{v} = -\zeta\grad{p}$ with a bulk mobility $\zeta$.
Note that as a gradient of a scalar pressure field, these velocity fields are always curl-free, thus the vorticity $\tens{\boldsymbol\omega}$ vanishes in this approximation.

With this approximation, the form of the expansion field $\vec{v}$ can now be directly computed from the local growth rate, the system geometry and the boundary conditions.
Together with the assumption that at low alignment strength, the nematic director closely follows the shear rate tensor $\ust$, we can thus predict steady-state alignment patterns for arbitrary geometries by solving a simple partial differential equation.

\begin{figure}
\centering
\includegraphics{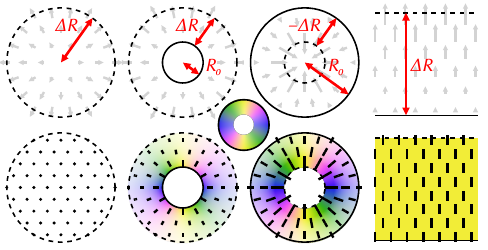}
\caption{Illustrations of the system geometry, isotropic-growth flow field (top row) and shear rate tensor (bottom row) for radial system geometry, with (from left to right) $R_0 = 0$ and $\Delta{}R>0$, $R_0>0$ and $\Delta{}R>0$, $R_0 > 0$ and $\Delta{}R < 0$, as well as the channel limit $R_0\rightarrow\infty$. Closed and open boundaries shown as solid and dashed black lines respectively. Colors and black bars in the bottom row represent orientation of the principal shear axis as indicated in the color legend, bar length and saturation represent shear strength.}
\label{pan:system-sketches}
\end{figure}

We apply this model to a generic family of systems: radially symmetric domains with mixed boundaries.
We impose a zero-velocity (i.e., impermeable) boundary condition at radius $R_0$, and a zero-pressure (i.e., open) boundary at $R_1 = R_0 + \Delta{}R$, yielding a ring-shaped domain with width $\Delta R$.
This description is a generalization of four geometries previously described in literature, as illustrated in Fig~\ref{pan:system-sketches}: with $R_0 = 0$, we reproduce the free radial expansion described by Dell'Arciprete et al.\cs\cite{dellarciprete_growing_2018, langeslay_strain_2024}.
Finite non-zero $R_0$ and positive $\Delta{}R$ corresponds to the systems with central non-growing regions as covered by Nijjer et al.\cs\cite{nijjer_mechanical_2021}, while the limit of $R_0\rightarrow\infty$ at constant $\Delta{}R$ approaches the commonly studied case of a rectangular channel or trap of length $\Delta{}R$\cs\cite{volfson_biomechanical_2008, orozco-fuentes_order_2013, isensee_stress_2022, langeslay_strain_2024}.
Finally, non-zero $R_0$ and negative $\Delta{}R$ yields flows reminiscent of the hole-filling inward growth described by Basaran et al.\cs\cite{basaran_largescale_2022, langeslay_strain_2024}.

Here, due to symmetry, both pressure and velocity can be described as scalar functions of radius only, thus converting equations \eqref{eq:velocity_divergence} and \eqref{eq:pressure_poisson} into one-dimensional ODEs, which are solved generally by
\begin{align}
p(r) &= \frac{\alpha}{4\zeta} \left[(R_0 + \Delta{}R)^2 + 2R_0^2\log(\frac{r}{R_0 + \Delta{}R}) - r^2\right]\label{eq:p_rad_fixedboundary}\\
\vec{v}(\vec{r}) &= \left(1-\Rr{2}\right)\frac{\alpha\,\vec{r}}{2}\label{eq:vel_rad_fixedboundary}\,.
\end{align}
We note that in this description, the velocity profile is independent of $\Delta{}R$, as---in the incompressible limit---the additional forces required to push the material downstream are always balanced by steric repulsion.
In the supplementary material, we show that this solution does indeed yield the known results for free radial and channel flow in the limits given above.

Taking the gradient, we arrive at the traceless symmetric shear tensor $\ust$ of this flow field:
\begin{equation}
\ust = \alpha\Rr{2}\left(\uvec{r}\otimes\uvec{r} - \frac{\tens{I}}{2}\right)\,,\label{eq:shear_radial_fixedboundary}
\end{equation}
where $\hat{r}$ is the radial unit vector and $\tens I$ is the unit tensor.
The principal direction and magnitude of this shear tensor are shown as color-coded plots for all four geometries in Fig.~\ref{pan:system-sketches}.

Assuming again that $\tens Q$ is approximately proportional to $\ust$, we therefore expect orientation along the radial direction, with alignment strength decaying with $r^{-2}$.
In the limit $R_0\rightarrow0$, $\ust$ vanishes and the shear-free circular case is restored.
In the channel limit $R_0\rightarrow\infty$, $\ust$ is constant in space as $r/R_0\rightarrow1$ for $R_0 \le r \le R_0 + \Delta{}R$.
We can thus qualitatively recover the expected results for all four cases: in the free radial flow, shear vanishes and we expect isotropy, while in all other cases, a non-vanishing shear flow acts as a source of radial orientational order.

\subsection{Agent-based numerical model}
\begin{figure}[t]
\includegraphics{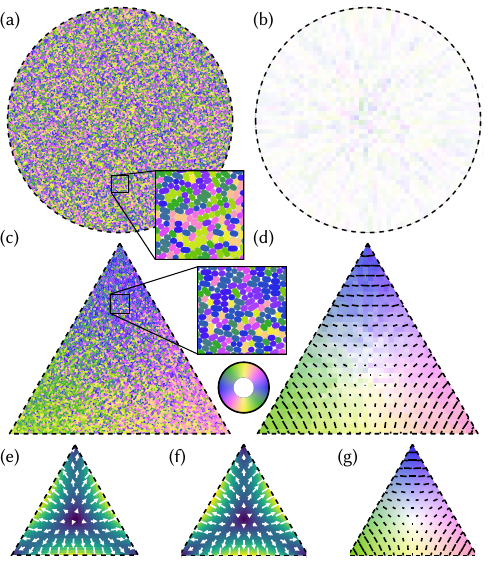}
\caption{
\panel{pan:bigcircle-snap}~Simulation snapshot of a circular colony of radius 100, containing approximately 23100 cells. Cells are color-coded by orientation. Inset shows a $15\times15$ section of the colony as indicated.
\panel{pan:bigcircle-field}~Orientation field, time-averaged over 10 generations, for the same simulation, using the same color code as for cells to represent the preferred orientation or nematic director $\vec{n}$. Saturation corresponds to the local nematic order parameter $q$ representing alignment strength.
\panel{pan:bigtriangle-snap},~\panel{pan:bigtriangle-field} Same as panels~\ref{onlyletter:pan:bigcircle-snap} and \ref{onlyletter:pan:bigcircle-field} for a triangular system with circumscribed radius of $100$, containing approximately 9400 cells.
\panel{pan:triangle-numericalvel} Time-averaged velocity field for the simulation in panel~\ref{onlyletter:pan:bigtriangle-snap}.
\panel{pan:triangle-theoryvel} Isotropic-growth flow field for a triangular domain.
\panel{pan:triangle-theoryshear} Principal direction of the shear rate tensor for the flow in panel~\ref{onlyletter:pan:triangle-theoryvel}.
}
\end{figure}
To test whether the isotropic-growth approximation also holds in less symmetric geometries with non-trivial director fields, we now introduce agent-based numerical simulations that allow us to validate our hypothesis beyond the geometries covered by existing literature.
We use a simple model introduced in previous work\cs\cite{isensee_stress_2022, hupe_minimal_2024} and inspired by similar models in literature\cs\cite{cho_selforganization_2007, volfson_biomechanical_2008, you_geometry_2018, orozco-fuentes_order_2013, vanholthetotechten_defect_2020, storck_variable_2014, langeslay_microdomains_2023, langeslay_stress_2024}.
Cells are modeled as rectangular rods of constant width (which, for simplicity, is set to $1$) with two semicircular caps.
During their life, cells grow from their original length $\lmax/2$ to $\lmax$ (both including the caps), then divide into two equally sized daughter cells.
The speed of this growth process is controlled by the particle doubling rate, here chosen randomly from the interval $\left[\frac{3}{4}\gamma_0, \frac{5}{4}\gamma_0\right]$, with the mean particle doubling rate $\gamma_0$ set to one.
This randomization of growth rates helps to desynchronize division processes across the colony.

Particles move according to overdamped equations of motion with simple mobilities that model sliding friction with a substrate, and interact only sterically, using a simple force law inspired by Hertzian repulsion.
Details on parameters and implementation of the model can be found in the supplementary material and previous work\cs\cite{hupe_minimal_2024}.

To simulate the impermeable zero-velocity boundaries introduced for the continuum model, we add immovable walls that interact sterically with the cells.
In the case of open zero-pressure boundaries, we remove cells from the system instantaneously when their center of mass position crosses the boundary.

The agent-based model, simulation and analysis software are implemented in the Julia programming language\cs\cite{bezanson_julia_2017}, using the open-source simulation framework InPartS\cs\cite{hupe_inparts_2022}. Unless specified otherwise, we use a maximum aspect ratio $\lmax=2$ for our rod-shaped cells, meaning that cells are initialized with a circular shape and extend along their backbone axis to reach an end-to-end length of $2$ units.
As a consistency check, Figure~\ref{pan:bigcircle-snap} shows a snapshot of a simulation of free radial expansion, in a circular domain with a radius of 100 cell diameters.
As expected, we find that in these systems orientations appear random with no apparent preferred orientation.
Due to the short division length of the cells, we do not observe large-scale microdomains as reported in other studies\cs\cite{you_geometry_2018, langeslay_microdomains_2023, isensee_sensitive_2025}.

To establish that orientations are indeed isotropically distributed everywhere in the domain, we can now compute time-averaged orientational order fields, exploiting that our system is in a steady state.
For this, we compute local estimates of $\tens Q$ by binning cell directors in a coarse rectangular grid with a bin size of several cell diameters\cs\footnote{The binning procedure used is described in detail by Isensee et al.\cs\cite{isensee_stress_2022}.}.
The result yields both the nematic director $\vec{n}$ (preferred orientation) and the nematic order parameter $q$ (alignment strength). This coarse-grained orientation field is shown in Figure~\ref{pan:bigcircle-field}, confirming
$q\approx0$ everywhere in the domain.

\subsection{Open polygonal systems}

As examples for systems with non-trivial director fields which could potentially be realized experimentally (e.g., in microfluidic devices), we consider domains shaped like regular polygons, which can be constructed by arranging straight, absorbing boundaries.
Analogous to the calculations in Section~\ref{sec:radialsystems_theo}, we expect these shapes to introduce an alignment bias, while also (unlike rectangular channels) being incompatible with the nematic symmetry of the cells and thus not supporting a globally uniform preferred orientation.

Figure~\ref{pan:bigtriangle-snap} shows a simulation snapshot for an open equilateral triangle with an circumscribed radius of 100 units, with the corresponding time-averaged orientation field in Figure~\ref{pan:bigtriangle-field}.
Local alignment strength is non-zero in large parts of the domain, being strongest at the outer edges, with preferred direction varying between the different regions.
At each corner of the domain, the orientation rotates by $-\pi/3$ in the lab frame, or by $-\pi$ relative to the radial direction, with cells being preferentially aligned tangentially close to the corners, and radially in between them.
This pattern can be quantified with the total topological charge of the orientation field, i.e.\ the number of rotations of the director along the boundaries of the system.
Here, this total charge is $-1/2$.
Towards the inside of the system, alignment strength decreases, with a relatively isotropic region located around the center of the domain.

Equation~\eqref{eq:pressure_poisson} can be solved analytically on equilateral triangles\cs\cite{volkov_property_1999} to obtain a closed-form expression for the velocity field (see supplementary material).
As a point of comparison, we compute the velocity profile of our agent-based simulations numerically, again averaging over snapshots in steady state.
Figures~\ref{pan:triangle-numericalvel} and \ref{pan:triangle-theoryvel} illustrate both velocity fields, showing a good qualitative match between theory and simulation.

With the traceless shear tensor $\ust$ of the theoretical velocity field, we can compute a pseudo-orientation field by interpreting $\ust$ as a nematic tensor.
We see that this pseudo-orientation, plotted in Fig.~\ref{pan:triangle-theoryshear}, successfully reproduces the orientation pattern observed in the agent-based simulations (Fig.~\ref{pan:bigtriangle-field}).

Using numerical solutions to the Poisson problem \eqref{eq:pressure_poisson}, we can extend this method to polygons with more than three sides (\suppfigFEFields) and find that the director rotates around each corner of the domain, analogous to the triangular case.
We can thus describe the expected behavior of the orientation fields in terms of the total topological charge.
In the observed pattern, we expect the nematic director to rotate clockwise by the interior angle of the polygon for every corner.
Thus, the expected defect charge is
\begin{equation}
\label{eq:total_topcharge}
s = 1 - \frac{n}{2}
\end{equation}
for an $n$-gonal domain.
Numerically, this can be computed by integrating orientation changes in the coarse-grained orientation field along a path parallel to the system boundaries.

In Fig.~\ref{pan:defect-sides}, we compare this theoretical prediction to defect charges obtained from the time-averaged orientation fields of agent-based simulations, on polygons ranging from $n = 3$ to $n = 17$.
For cells of division aspect ratio 2, the computed charge follows this prediction exactly for the entire range of $n$.
If we instead use instantaneous orientation fields without time averaging, the defect charge starts fluctuating upwards of the expected value for $n \ge 5$, with the mean also drifting up with increasing $n$.

For more elongated cells with division aspect ratio $\lmax=3$, the general behavior is similar; however here, the measured charge for the time-averaged field departs from the prediction at $n=12$, while the instantaneous charges already deviate for square domains.
This is indicative of the coarser microstructure of colonies with more anisotropic cells (compare snapshots in \suppfigPolySnaps) which induces larger fluctuations.

\begin{figure}[ht]
\includegraphics{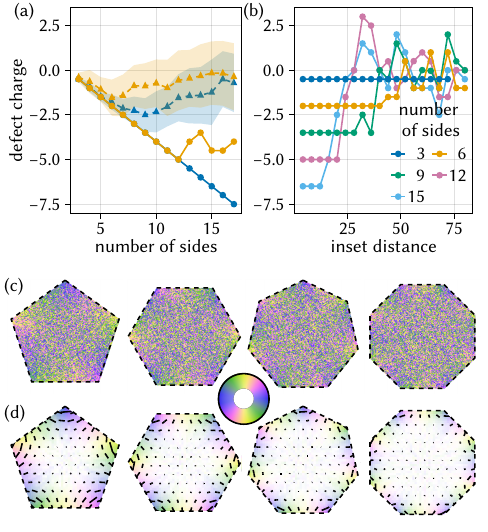}
\caption{
\panel{pan:defect-sides}~Defect charge in polygonal domains as a function of number of sides, computed on a time-averaged orientation field (circles, solid lines) as well as by averaging charges from instantaneous orientation fields over snapshots (triangles, dashed lines), for cells of division aspect ratio 2 (blue) and 3 (orange). Shading indicates standard deviation over snapshots.
\panel{pan:defect-inset}~Total defect charges of the time-averaged orientation fields on various polygonal domains, computed using a path at varying inset distances from the outer domain boundary, for cells with division aspect ratio $2$.
\panel{pan:polys-snap}~Snapshots of simulations of 5-, 6-, 7- and 8-gonal domains with an circumscribed radius of $100$. Cells are color-coded by orientation.
\panel{pan:polys-field}~Spatially coarse-grained time-averaged orientation fields for the same simulations, with color saturation corresponding to alignment strength in the time average.
}
\end{figure}

We can also use the topological charge to resolve the spatial variation in alignment strength by changing the inset distance between the integration path and the boundary.
Fig.~\ref{pan:defect-inset} shows that the defect charge generally remains constant before transitioning to a fluctuation around zero beyond some critical distance.
This corresponds to the region of low order observed qualitatively in the orientation fields.

We also see that the outer region in which the charge remains constant decreases in width with increasing number of corners.
This result shows how the prediction of linearly decreasing topological charge is consistent with the limit of $n\rightarrow\infty$.
As there is no persistent alignment in the circular domain, we expect the topological charge to vanish, while the empirical law in Eq.~\eqref{eq:total_topcharge} would predict diverging topological charge.
However, as the number of sides increases, the depth of the well-ordered outer layer in which the law applies decreases, while the radius of the near-isotropic interior with zero defect charge increases.
Thus, in the circular limit, the outer layer has vanished and only the disordered interior remains.

In summary, we find that even in geometries with more complex variations of the nematic director, the isotropic-growth approximation can be used to obtain a good qualitative prediction for orientations and alignment strengths from the shear rate tensor.
One down-side of this approach is that $\ust$ itself does not set a scale for $\tens Q$. Therefore, it would be desirable to extend this approach in a minimal way that allows for a quantitative prediction. To do this, we turn back to the radial systems introduced in Section~\ref{sec:radialsystems_theo}, as they are better-suited for a systematic comparison between predicted and numerical values due to their symmetry properties.

\section{Quantitative analysis: Advection-decay model}

\subsection{Motivation and setup}
\label{sec:radialsystems_num}

Out of the four geometries that can be realized with the description introduced in Section~\ref{sec:radialsystems_theo}, we choose the case of non-zero $R_0$ and positive $\Delta{}R$, i.e.\ a disk-shaped impermeable obstacle of radius $R_0$ surrounded by an absorbing ring at $R_0 + \Delta{}R$ (Fig.~\ref{pan:ring-snap}).
These ring-shaped systems show high spatial variation of alignment strength, while being numerically more accessible close to the incompressible limit than systems with inward flow due to the lower stresses and velocities at equivalent growth rate.

We simulate colonies in domains with varying $R_0$, keeping $\Delta{}R$ constant.
In order to speed up computation and access larger $R_0$ without increasing the number of particles in the system substantially, we use wedge-shaped sections of approximately constant area instead of full rings.
The sides of these wedges are closed with straight walls aligned radially, conforming to the flow field in a full ring.
This also allows to take the ``channel limit'' $R_0\rightarrow\infty$ as described in Section~\ref{sec:radialsystems_theo} in our agent-based simulations (see Fig.~\ref{pan:channel-snap}).
For analysis of cell orientations, we exclude the regions close to these walls to eliminate the influence of potential boundary effects.

We compute a radial order parameter $q_{rad}$ as a function of radial position, defined as
\begin{equation}
\label{eq:q_rad}
q_{\text{rad}} = 2\langle(\uvec{n}_{\text{cell}} \cdot \uvec r)^2\rangle - 1\,,
\end{equation}
where $\uvec r$ is the radial unit vector at the cell position and the average is over all cells in an annular bin.
Assuming that $\tens Q$ is of the form $q(r) (\uvec r\otimes\uvec r - \tens I/2)$, as required by the symmetry of the system, this order parameter $q_\text{rad}$ is equal to $q(r)$ (see supplementary material).
Thus, should the proportionality of $\tens Q$ and $\ust$ hold, we expect it to decay with $r^{-2}$ according to Eq.~\eqref{eq:shear_radial_fixedboundary}, although we cannot predict its absolute scale.
However, Figure~\ref{pan:ring-alignment} shows that this scaling is not accurate:
While the measured order parameter decay follows a power law, it decays significantly slower than expected, with decay exponent varying with decreasing with increasing $R_0$.
This result shows that for short cells, local shear anisotropy alone cannot quantitatively predict the alignment strength.

\begin{figure*}[ht]
\includegraphics{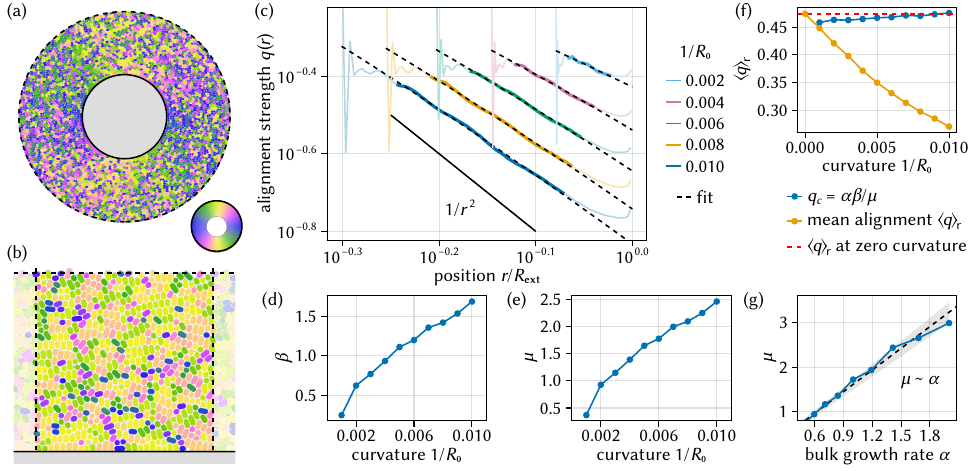}
\caption{
\panel{pan:ring-snap}~Snapshot of a colony in a ring-shaped domain with $R_0 = 20$ and $\Delta{}R = 30$. Cells are color-coded by orientation as indicated by the color legend.
\panel{pan:channel-snap}~Snapshot of a colony in an $x$-periodic channel with $\Delta{}R = 30$, with a closed lower boundary and an open upper boundary. Cells are color-coded as in panel \ref{onlyletter:pan:ring-snap}.
\panel{pan:ring-alignment}~Mean alignment strength $q$ in ring-shaped domains of constant height $h=100$ and varying inner radius $R_0$ with fits of the advection-decay solution (dashed lines).
\panel{pan:ring-fit-beta}~and~\panel{pan:ring-fit-mu}~Fit parameters $\beta$ and $\mu$ of the advection-decay solutions.
\panel{pan:ring-fit-qnull}~Comparison of $q_c=\alpha\beta/\mu$ computed from the advection-decay fit parameters (blue circles) and the mean alignment strength (orange circles) in rings of varying inner radius $R_0$.
\panel{pan:ring-growthrate}~Fitted decay rate $\mu$ in a ring of inner radius $R_0 = 200$ and height $h = 100$ over bulk growth rate $\alpha$. For all simulations in panels \ref{onlyletter:pan:ring-alignment} to \ref{onlyletter:pan:ring-growthrate}, for computational efficiency, wedges (i.e., sectors of the full annulus) were simulated while keeping the total area filled constant at 10000 area units, corresponding to a $100\times 100$ channel.
}
\end{figure*}

By assuming that the mean orientation is always aligned with the local principal axis of the shear flow, we have implicitly assumed that a cell, advected in the colony expansion flow, can adapt its orientation to follow changes of the shear instantaneously.
However, physically, we expect the process of reorientation to take a certain amount of time, limited e.g.\ by the cell mobility and the local microstructure of the colony: if a patch of highly aligned cells is advected into a region of low shear and becomes unstable, it will not instantly become isotropic.
When this finite timescale of particle rotation becomes relevant, an advection term accounting for the flow velocities in the growing colony should therefore be considered for the  $\tens Q$ description.

We turn back to Eq.~\eqref{eq:basic_q} to derive an approximation for the steady state $\tens Q$ that includes advection effects, while keeping the isotropic-growth assumption and also keeping the model analytically tractable for radially symmetric domains.
We set the temporal derivative of $\tens Q$ to zero, leaving only an advective spatial derivative.
On the right hand side, we keep the shear alignment term, scaled by an alignment strength parameter $\beta$.
As we assume the velocity field to be well approximated by the gradient of a scalar pressure, we can neglect the rotational terms.
Finally, to compensate for the constant source term, we introduce a simple linear decay with rate $\mu$, resulting in
\begin{equation}
(\vec v\cdot\nabla)\,\tens Q = \beta\ust - \mu\tens Q\,.\label{eq:adv_dec}
\end{equation}
This constant source term is the simplest possible term that ensures that $\tens Q$ vanishes in the absence of shear, which is required for a description of the steady state.

Using the radial symmetry of the system, we can simplify this equation, as we know that both the shear rate tensor and nematic tensor must be of the form $f(r) (\uvec r\otimes\uvec r - \tens I/2)$ with a scalar function $f$.
Inserting this into the advection-decay equation, we obtain the one-dimensional ODE
\begin{equation}
v(r) \pdv{r} q(r) = \beta u^{ST}(r) - \mu q(r)\,,
\end{equation}
which is solved by the expression
\begin{equation}
\label{eq:adv_fixedboundary}
q(r) =\, q_c  \, \hypergeo{r} + C \alpha \left(r^2 - R_0^2\right)^{-\frac{\mu}{\alpha}}
\end{equation}
where $_2F_1$ is the Gaussian hypergeometric function, $q_c = \alpha\beta/\mu$ and $C$ is a constant of integration.
These two terms correspond to different processes: the first term, arising from the inhomogeneity of Eq.~\eqref{eq:adv_dec}, describes the alignment created by shear within the colony, while the second term (the solution of the homogeneous equation) corresponds to the decay of alignment injected at the boundary.
Assuming that $\tens Q$ does not diverge at $R_0$, we set the integration parameter $C$ to zero to obtain a consistent solution.

Without the homogeneous decay term, the magnitude of $\tens Q$ at $r = R_0$ is now entirely determined by the system parameters:
Inserting $r = R_0$ into Eq.~\eqref{eq:adv_fixedboundary}, we find
\begin{equation}
q(r = R_0)  = q_c = \frac{\alpha \beta}{\mu}\,.
\end{equation}
This constant is of particular interest, as taking the limit $R_0\rightarrow\infty$ with $r/R_0\rightarrow 0$ yields the same value as a prediction for the alignment strength in a channel.

The decay exponent of the advection-decay solution is no longer constant in space, but for radii close to $R_0$ it can be determined analytically by transforming to logarithmic coordinates $x = \log(r/R_0)$ and expanding $\log(q(x))$ to first order around $x = 0$, yielding
\begin{equation}
\label{eq:advdec_exponent}
\log(q(r)) = \log(q_c) - \frac{2\textstyle{\frac{\mu}{\alpha}}}{1 + \textstyle{\frac{\mu}{\alpha}}}\log(r/R_0) + \mathcal{O}(\log(r/R_0)^2)\,.
\end{equation}
This shows that the exponent of the alignment profile is set by the decay parameter $\mu$: As alignment decay weakens ($\mu\rightarrow0$), the profile becomes flatter as more aligned material is advected from regions of higher shear.
If decay is fast and $\mu$ large, the alignment strength exponent converges to two, thus following the exponent of the shear profile.

\subsection{Parameter fits}

Figure~\ref{pan:ring-alignment} shows fits of the advection-decay solution to the radial order parameter profiles obtained from our agent-based simulations.
We find that in the range of near-power law decay close to the central obstacle, the model fits the numerical data well, although the numerical curve deviates from the fit close to the outer edge of the domain.
We compute the parameter $q_c$ from the fit parameters, as well as measuring it directly from the average alignment strength in a channel of corresponding dimensions.
As shown in Fig.~\ref{pan:ring-fit-qnull}, we find the values to be in good agreement, although the $q_c$ computed from the fit parameters is lower than the channel limit for low curvature, and appears to increase slightly with increasing curvature.

The individual fit parameters $\beta$ and $\mu$ (Fig.~\ref{pan:ring-fit-beta}-\ref{onlyletter:pan:ring-fit-mu} are less well-behaved:
As the alignment profiles get slightly steeper with increasing curvature, while the alignment strength at $R_0$ stays relatively constant, both $\mu$ and $\beta$ increase with $1/R_0$.
Approaching the limit of $1/R_0 \rightarrow 0$, the fit parameters appear to vanish.
However, in this limit, the fit is ill-conditioned, as the advection-decay solution here only depends on the ratio $q_c$ and becomes constant in space, thus making separate inference of $\mu$ from the decay exponent impossible.
An investigation of the mean squared error as a function of both parameters (\suppfigFitLandscape) reveals a degenerate fit landscape even at intermediate curvatures, with fit error mostly controlled by the ratio of $\mu$ and $\beta$.

Looking at the advection-decay solution, we see that the decay rate $\mu$ only appears relative to the growth rate $\alpha$.
In the incompressible limit, steric interactions respond instantly to strain caused by cell growth.
Thus, we expect the steady-state alignment profile to be independent of growth rate in this limit.
Therefore, $\mu$ should be proportional to $\alpha$ to keep the alignment profile unaffected.
Numerically, we confirm this by varying the division rate of the particle while keeping other mechanical properties constant.
Fig.~\ref{pan:ring-growthrate} shows that the fit parameter $\mu$ scales with growth rate (Fig.~\ref{pan:ring-growthrate}), indicating that the decay of order is ultimately also driven by the central active process in the system, namely proliferation.

In summary, we see that the simple advection-decay model yields quantitatively accurate order profiles and therefore correctly accounts for the additional time scales involved in translating the shear rate profile in ring-shaped systems into radial alignment through reorientation and advection.

\subsection{Tangential alignment: Excess growth}
\begin{figure*}[ht]
\includegraphics{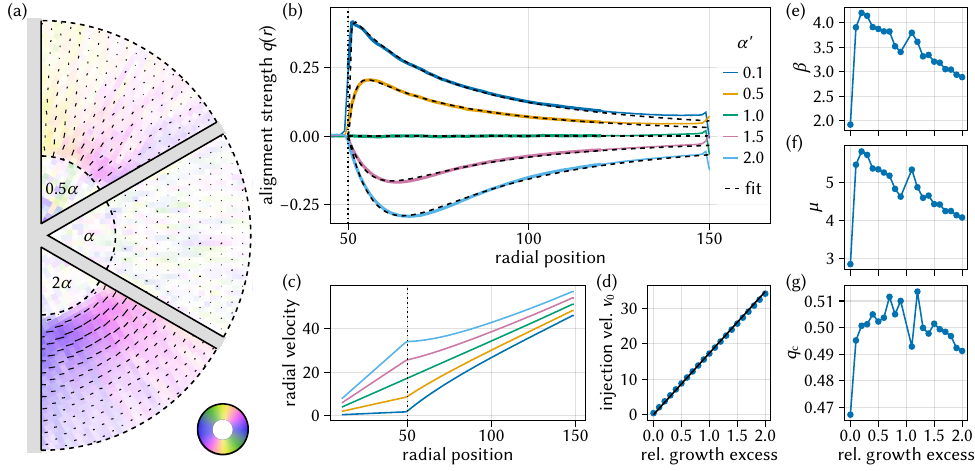}
\caption{
\panel{pan:excess-pie}~Time-averaged orientation fields for wedge-shaped domains with $R_0 = 50$ and $\Delta{}R = 100$, with modified growth rate $\alpha^\prime$ for $r < R_0$, and (from top to bottom) $\alpha^\prime = 0.5\alpha$ resulting in radial alignment, $\alpha^\prime = \alpha$ resulting in no alignment and $\alpha^\prime=2\alpha$ resulting in tangential alignment.
\panel{pan:excess-alignment}~Mean alignment strength $q$ in wedge-shaped with varying $\alpha^\prime$, with fit of the expanded advection-decay solution.
\panel{pan:excess-velocity}~Mean radial velocity for the same systems as panel~\ref{onlyletter:pan:excess-alignment}.
\panel{pan:excess-vnull}~Numerical velocity at $r = R_0$ as a function of growth excess $\alpha^\prime/\alpha$ for the systems in panel~\ref{onlyletter:pan:excess-alignment} and \ref{onlyletter:pan:excess-velocity}, compared with the analytical prediction (black line).
\panel{pan:excess-fit-beta}, \panel{pan:excess-fit-mu} and \panel{pan:excess-fit-qnull}: Fit parameters $\beta$ and $\mu$, and $q_c = \alpha\beta/\mu$ for the advection-decay fits in panel~\ref{onlyletter:pan:excess-alignment} as a function of growth excess $\alpha^\prime/\alpha$.
}
\end{figure*}

To further test the performance of the model, we now create a system that also features tangential alignment, corresponding to $q < 0$.
We can deliberately construct such a system by setting the appropriate shear anisotropy: if a lower velocity divergence in the center of the domain gives rise to a flow with radially biased shear and thus induces radial alignment, a higher velocity divergence should change the sign of the shear tensor and create a tangentially biased shear.
Analytically, this can be represented by relaxing the zero-flow boundary condition at $R_0$ to allow a constant injection velocity $v_0$.
We can again solve the Poisson problem to derive a velocity field for the incompressible isotropic case in $d$ dimensions, yielding
\begin{equation}
\vec{v}(\vec{r}) = \left(\left(\frac{v_0}{R_0} - \frac{\alpha}{d}\right)\,\left(\frac{R_0}{r}\right)^d + \frac{\alpha}{d}\right)\,\vec{r}\,.\label{eq:flow_radial_v0}
\end{equation}
and a traceless symmetric shear tensor $\ust$ of
\begin{equation}
\ust = d\left(\frac{\alpha}{d} - \frac{v_0}{R_0}\right)\left(\frac{R_0}{r}\right)^d\left(\uvec{r}\otimes\uvec{r} - \frac{\tens{I}}{d}\right)\,\label{eq:shear_radial_v0}
\end{equation}
We can see from this result that $\ust$ becomes zero at $v_0/R_0 = \alpha/d$, corresponding again to shear-free expansion, and changes direction for higher $v_0$.
Thus, for a large enough injection velocity, we expect alignment to become tangential instead of radial.

To confirm this prediction with our agent-based model, we use a simple wedge-shaped domain with spatially modulated particle growth rate.
For particles within $r < R_0$, we set the growth rate to $\alpha^\prime =  f\,\alpha$ to control the mean radial velocity at $R_0$.
Figure~\ref{pan:excess-pie} shows system snapshots for three example systems, with weak radial alignment for $\alpha' = 0.5\alpha$, no alignment for $\alpha' = \alpha$, and tangential alignment for $\alpha' = 2\alpha$.

Using the generalized radial flow solution, we can adapt the advection-decay model to these systems, yielding a solution of similar form to Eq.~\eqref{eq:adv_fixedboundary}.
The full analytical solution for these boundary conditions is shown in the supplementary material.

Here, fixing the homogeneous part of the solution requires imposing a boundary condition on the injected $q$ at the inner border for all $v_0 \neq 0$.
Using our observations that the inner sector of the domain behaves like a freely expanding system, with a linear velocity profile (Figure~\ref{pan:excess-velocity}) and no alignment bias (Figure~\ref{pan:excess-alignment}), we set $q(R_0) = 0$.
To fit the advection-decay solutions to the data using only the model parameters $\beta$ and $\mu$, we additionally need to obtain a value for the injection velocity $v_0$.
For this, we use the analytical expression for free expansion with growth rate $\alpha'$, which matches the measured velocities at $R_0$ (Figure~\ref{pan:excess-vnull}).

Figure~\ref{pan:excess-alignment} shows the advection-decay fits for different $\alpha'$.
As before, the fits provide a good match to the data, with minor deviations from the measured alignment strength towards the outer edge.
Similar to the ring systems, the fit parameters $\beta$ (Figure~\ref{pan:excess-fit-beta}) and $\mu$ (Figure~\ref{pan:excess-fit-mu}) vary with $\alpha'$, here decreasing with increasing injection velocity, while the ratio $q_c = \alpha\beta/\mu$ stays relatively constant at a value of around 0.5, as shown in Figure~\ref{pan:excess-fit-qnull}.

\section{Discussion}

We showed that in systems of growing rods, spatially heterogeneous average alignment patterns can be qualitatively predicted from domain geometry only.
For this, we assume that the preferred orientation of rods is determined by the shear rate of the expansion flow, analogous to classical nematics.
We break the feedback between orientation dynamics and expansion flow by assuming that for sufficiently short cells, the flow field generated by anisotropic growth is similar to that of isotropic growth, thus enabling us to obtain simple estimates for the shear rate tensor using only the cell growth rate and the boundary conditions of the flow field.
Interestingly, we find that, in the time average, the isotropic-growth approximation holds well even for more anisotropic cells with more pronounced local nematic alignment, although instantaneous fluctuations naturally increase.

While the authors of Ref.\cs\citenum{you_confinementinduced_2021} argue that flow alignment alone, while sufficient to predict the preferred orientation in a channel, is not consistent with the absence of radial alignment in a freely expanding colony (because of the strong radial flow), our framework shows that we indeed expect no preferred alignment. This discrepancy in interpretation is rooted in the tracelessness of $\ust$, which essentially eliminates the \emph{isotropic} component $\div\vec{v}$ of the expansion and only retains the anisotropic component relative to it: While there clearly is a radial flow during free radial expansion, still $\ust=\tens 0$, i.e., the expanding material is distributed locally isotropically (which has been compared to Hubble expansion in Ref.\cs\citenum{dellarciprete_growing_2018}), resulting in no flow alignment. In contrast, any anisotropic \emph{redistribution} induced by the boundary conditions results in nonzero $\ust=\tens 0$ and thus alignment such as in the channel where all flow is redirected towards the outlets. The other considered geometries then show that this argument can be generalized to spatially varying $\ust$.

\begin{figure}[h]
\includegraphics{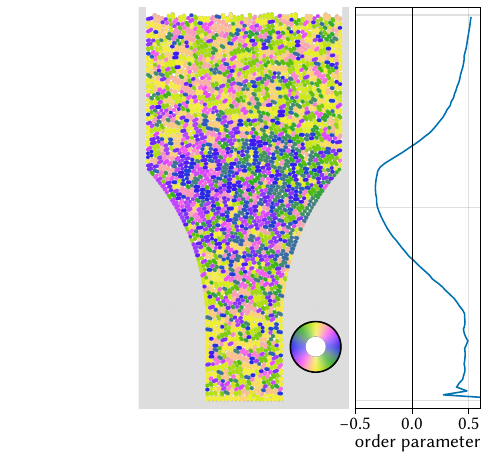}
\caption{
Snapshot of a channel-like system engineered to rotate preferred orientations from vertical to horizontal and back to vertical. Right-hand panel shows the scalar product order parameter analogous to Eq.~\eqref{eq:q_rad}.
\label{fig:trumpet}
}
\end{figure}

Practically, our framework could be applied, e.g.\ in microfluidic experiments to engineer systems with desired alignment patterns, where the nematic orientation changes in a prescribed manner along the flow.
As a feasible demonstration, we modified a channel-type geometry to include a region where the channel width expands fast enough to flip the direction of the expected shear rate tensor.
In this region only, it will be oriented horizontally, analogous to the tangential alignment seen in the radially symmetric systems with excess growth, while in the channel-like regions surrounding it, the principal shear component will be vertical.
As seen in Figure~\ref{fig:trumpet}, particles in this system do indeed change orientation from vertical to horizontal while flowing through this region.
Orientation patterns in expanding materials based on precomputed shear-flow properties could therefore also be used to engineer specific rotational dynamics in addition to static properties.

In quantitative comparisons in systems with spatially varying alignment strength, we saw that the shear rate alone is not sufficient to describe the observed change accurately.
Based on an evolution equation for the nematic tensor, we therefore proposed a simple model for the steady state alignment profile, incorporating the effects of advection and decay of alignment.
For this, we chose to approximate the effects that decrease time-averaged alignment by a linear decay term, effectively balancing the non-local creation of order with a local decay.
Interestingly, fits to numerical simulations with varying growth rates indicated that the decay rate is proportional to the growth rate. This is noteworthy because it suggests that, in absence of thermal noise, fluctuations induced by growth and division themselves are the primary sources of orientational noise, which is consistent with prior studies that found an effective noise arising from local rearrangements\cs\cite{sunkel_motilityinduced_2025,lish_isovolumetric_2024}. With growth and division causing both the shear flow responsible for orientational order as well as the decay of order, the activity thus replaces both the free energy for spontaneous alignment as well as thermal fluctuation counteracting it, underscoring the manifestly non-equilibrium nature of the system.

This is in contrast to previous models for similar systems (e.g.\ \cite{volfson_biomechanical_2008}), where the production of $\tens Q$ was controlled by modulating the source terms of the $\tens Q$ evolution equation with a factor of $(1 - \norm{\tens Q})$.
As a consequence, in a channel flow, where shear is constant and homogeneous in space, these equations are only solved by states with perfect nematic order.
In simulations, these states are only observed for cells with aspect ratio larger than $\gtrapprox4$\cs\cite{isensee_stress_2022}, so these models do not apply to the shorter cells studied here.

In recent work, Langeslay and Juarez\cs\cite{langeslay_strain_2024} introduced a model for the alignment direction and strength, also using the shear component of the expansion flow as the principal source of order, but using a different argument:
Their model is derived from the basic assumption that in a growing monolayer, the global flow strain rate can be decomposed into particle-level contributions caused by rotation and expansion.
Using this ansatz, they make predictions for channels and free expansion, as well as for the case of a colony growing on a sphere from one pole, where they find a quantitative match between measured alignment strength and prediction.
However, they note that their assumption may not hold at lower nematic order, where local rearrangements of particles can also contribute to strain.
Arguing that for sufficiently elongated cells, these rearrangements are effectively prohibited, they thus conclude that their model serves only as an approximation for cells of high division aspect ratio.
Correspondingly, their model also predicts parameter-independent perfect alignment in channel geometries.

In contrast, the advection-decay model presented in this work is based on the assumption that growth is always close to isotropic, and that the creation of nematic order by non-isotropic strains is balanced by a decay term, primarily driven by growth- and division-induced fluctuations.
Microscopically, the latter assumption can only hold for low division aspect ratio: a spontaneous decay process that reduces order while conforming to the global expansion flow shear rate requires particle rearrangements.
Our model can thus be seen as a description for low aspect ratio rods, complementing previous work that focused on the high order limit.

While the advection decay model is able to reproduce the shape of alignment strength profiles in radial systems, its solutions depend on the values of the unknowns $\beta$ and $\mu$.
Using parameter fits at varying ring curvature, we find the best fit values for these parameters are not constants depending only on the properties of the expanding cell fluid, but change with ring geometry.
Similarly, in rings with excess growth in the center, the best fit values of $\beta$ and $\mu$ vary with the degree of growth excess.
This indicates that here, our description is too simplistic, and other systematic effects such as e.g.\ mechanical stresses and their gradients may contribute to the balance of alignment creation and decay.
As an example, in the supplementary information we show analytically that the pressure at the fixed boundary doubles between the circular and channel limits. However, it is important to note that, despite the geometry dependence of individual parameters, certain parameter combinations still allow for predictions across geometries, such as the ratio $q_c=\alpha\beta/\mu$ for the order parameter in a channel.

For all theoretical considerations in this work, we have assumed incompressibility of cells, and correspondingly constant density throughout the colony.
Numerically, as cell hardness is finite and mechanical stresses increases towards the center of the colony (compare e.g.\ Eq.~\eqref{eq:p_rad_fixedboundary}), incompressibility can only be approximate.
As we expect density profiles to be parabolic, we would expect the largest deviations from the expected velocity profiles to occur towards the outside of the domain, where density (and thus effective growth rate) decays fastest.
By extending Eq.~\eqref{eq:velocity_divergence} to allow for a static, non-uniform density profile similar to that measured in the simulations, we can show that in this case, the shear increases towards the outside of the domain (\suppfigHockeyStick).
In the alignment data (Figures~\ref{pan:ring-alignment} and \ref{pan:excess-alignment}) we do indeed observe a matching increase in shear rate $\ust$ for high radii.
As the increased complexity of the equations makes analytically solving the advection-decay model with a non-uniform density field infeasible, we instead restrict the fit range for the advection decay parameter estimation to the inner regions of the domain, where density gradients are lower and expected deviations smaller.

In conclusion, the isotropic growth approximation enables the simple qualitative prediction of alignment patterns in dense colonies of growing rods and thus the design of geometries that yield desired alignment patterns. Using the same information, i.e.\ without any feedback of active stresses on the velocity field, the advection-decay model can be used to refine the approximation and yield quantitatively accurate alignment strengths. Based on the dominant influence of system topology on the resulting alignment patterns, these mechanisms could also contribute to the interpretation of cell arrangements in fast-growing biological tissues and open the door towards the design of more complex arrangements of topological defects in artificial materials designed through volume growth.

\section*{Code availability}
Simulations used for this study were built in Julia\cs\cite{bezanson_julia_2017}, using the open-source simulation framework \InPartS{}\cs\cite{hupe_inparts_2022}. The implementation of the agent-based models is part of the \InPartS{}Biome model library, published at \url{http://hdl.handle.net/21.11101/0000-0007-FE13-6}.
Example code for simulation and data analysis will be made available alongside the final publication.
Visualizations were prepared using Makie.jl\cs\cite{danisch_makie_2021}.

\begin{acknowledgments}
We acknowledge support from the Max Planck School Matter to Life and the MaxSynBio Consortium, which are jointly funded by the Federal Ministry of Education and Research (BMBF) of Germany and the Max Planck Society.
\end{acknowledgments}

\end{document}